%% file: mssm_sent_to_prl_v10.tex
\documentstyle[eqsecnum,epsf,aps,prl,preprint]{revtex}
\input{psfig}

\def\met{\mbox{$\not\!\!{E}_{T}\,$}}
\def\et{\mbox{$E_T$}}
\def\pt{\mbox{$p_T$}}
\def\tb{\mbox{$\tan\beta$}}

\begin{document}
\onecolumn
\draft


\vspace*{0.2in}
\begin{center}
\begin{large}
\bf Search for Neutral Supersymmetric Higgs Bosons \\ 
in $p\bar{p}$ Collisions at $\sqrt{s}=1.8$ TeV
\end{large}
\end{center}

\normalsize
\vspace*{1.5cm}
\begin{center}
{\large The CDF Collaboration} 
\end{center}

\vspace*{0.5cm}

\begin{abstract}
\baselineskip 24pt
We present the results of a search for neutral Higgs bosons produced
in a\-sso\-cia\-tion with $b$ quarks in $p\bar{p}\rightarrow b\bar{b}
\varphi\rightarrow b\bar{b}b\bar{b}$ final states with
$91 \pm 7$ pb$^{-1}$ of $p\bar{p}$ collisions at $\sqrt{s}=1.8$ TeV 
recorded by the Collider Detector at Fermilab. 
We find no evidence of such a signal and the data is interpreted
in the context of the neutral Higgs sector of the Minimal Supersymmetric 
extension of the Standard Model. With basic parameter choices for the 
supersymmetric scale and the stop quark mixing, we derive 95\% C.L. 
lower mass limits for neutral Higgs bosons for $\tb$ values in excess 
of 35.
\end{abstract}

\vspace*{0.2cm}

\maketitle
\input{authors.tex}

\newpage

\onecolumn

\baselineskip 24pt
A fundamental question which remains open today in particle 
physics is the origin of the electroweak symmetry breaking.
The simplest mechanism in the Standard Model (SM) and in many supersymmetric 
extensions is spontaneous symmetry breaking, achieved
through the introduction of one or more scalar field doublets. 
The SM assumes one doublet of scalar fields and a single physical 
Higgs boson ($h_{SM}$), with unknown mass but with fixed couplings 
to other particles.
A more complex symmetry breaking mechanism occurs in the Minimal
Supersymmetric extension of the Standard Model (MSSM), where several 
physical scalar states are predicted: three neutral bosons 
(the CP-even $h$ and $H$, and the CP-odd $A$) and two charged 
bosons ($H^{\pm}$). A distinct feature of the MSSM is the modified 
couplings of the Higgs particles,
in particular the enhancement of the bottom-Higgs Yukawa couplings by
$\tb$ for the case of the $bbA$ vertex, where $\tb$ is the ratio of 
the vacuum expectation values of the two Higgs doublets of the theory.
The Higgs sector of the MSSM is completely determined at tree level by 
two free parameters, chosen to be the mass of the CP-odd Higgs boson, 
$m_A$, and $\tb$. The mass of the CP-even Higgs boson $h$, $m_h$,
is constrained to be less than $m_{Z^0} |\cos 2\beta|$. Radiative corrections 
substantially modify the masses and couplings of the two neutral 
CP-even Higgs scalars, in particular the upper bound on $m_h$~\cite{fin}.
\par
At the Tevatron, one of the Higgs production mechanisms likely
to be observed in both the SM and some regions of the MSSM parameter space 
is the associated production
$p\bar{p}\rightarrow V + \varphi$, where $V = W, Z$
and $\varphi = h, H,h_{SM}$. CDF has already reported on searches for this 
channel ($V + h_{SM}$) with different signatures~\cite{cdflep,cdfhad}. 
In this Letter we exploit the enhanced bottom-Higgs Yukawa couplings 
of the MSSM to test the large $\tb$ sector of the theory 
by searching for the process
$p\bar{p}\rightarrow b\bar{b}\varphi \rightarrow b\bar{b}b\bar{b}$
with $\varphi = h,H,A$. 
Our sensitivity in this search is limited to the region of parameter space
corresponding to $\tb \gtrsim 35$. In this region, at least one of the CP-even 
Higgs bosons has similar couplings to, and is always degenerate with, the 
CP-odd Higgs boson. For $m_A\lesssim 110-125$ GeV/c$^2$ 
(depending on $\tb$ and on the 
parameters of the stop quark mass matrix), $m_h \simeq m_A$, while
for $m_A\gtrsim 110-125$ GeV/c$^2$, $m_H \simeq m_A$. 
Therefore this analysis covers a simultaneous search for two or more Higgs 
signals with an experimental signature of four $b$ jets in the final state.
\par
The search reported here is based on $91 \pm 7$ pb$^{-1}$ of
integrated luminosity recorded
during the 1994-95 Tevatron run.
The CDF detector is described in detail elsewhere~\cite{cdf_d}. 
The silicon vertex detector
(SVX) consists of four layers of axial microstrips located 
immediately outside the beampipe with an innermost radius
of 2.9 cm~\cite{svx}. The SVX provides precise track 
reconstruction in the plane perpendicular to the beam and 
the ability to identify secondary vertices 
produced by heavy flavor decays. The momenta of charged particles
are measured in the central tracking chamber (CTC), which lies inside
a 1.4 T superconducting solenoidal magnet.
Outside the CTC, electromagnetic and
hadronic calorimeters arranged in a projective tower geometry cover the
pseudorapidity region $|\eta| < 4.2$~\cite{CDFcoo} and are used
to identify jets. The data sample was recorded with a trigger
which requires four or more clusters of contiguous 
calorimeter towers, each with transverse energy $E_T \ge 15$ GeV, 
and a total transverse energy $\sum E_T \ge 125$ GeV. 
\par
The initial steps of the data selection are the same as in the recent CDF
SM Higgs search~\cite{cdfhad}.
We start by rejecting cosmic ray events, beam halo, and detector noise. 
Events with one or more identified electrons or muons from vector 
boson decays defined as in~\cite{cdflep} are also rejected.
After this selection, events are required
to have at least four jets with $E_T\ge 15$ GeV and well contained 
within the fiducial calorimeter regions $|\eta|\le 1.5$. 
Jets are defined as localized 
energy depositions in the calorimeters and are reconstructed 
using an iterative clustering algorithm with a fixed cone of radius
$\Delta R = \sqrt{\Delta\eta^2 + \Delta\phi^2} = 0.4$ in 
$\eta - \phi$ space~\cite{jets}. Jet e\-ner\-gies 
are corrected for energy losses in uninstrumented detector regions, 
energy falling out\-side the clustering cone, contributions from underlying 
event and multiple interactions, and calorimeter nonlinearities.
The selected data sample is dominated by QCD multijet events and contains 
$207,604$ events.
The four highest-$E_T$ jets in an event are then ordered in $E_T$
and a search sample is obtained for each of a set of Higgs boson masses by
requiring the three highest-$E_T$ jets to pass cuts which are Higgs boson
mass dependent.
This is motivated by the fact that the $E_T$ spectrum of the leading jets 
for the signal is, on average, larger than the QCD background, and grows
with increasing scalar boson mass ($m_\varphi$).
We use the leading order (LO) parton level matrix elements~\cite{tim} 
encoded in the PYTHIA v5.6 Monte Carlo program~\cite{pythia} along with
a full simulation of the CDF detector to simulate the signal. 
We use the CTEQ3L parton distribution functions and set a factorization 
scale equal to the Higgs boson mass in the simulation.
We find optimal $E_T$ cuts by maximizing the expected 
significance of the signal. For a Higgs boson mass of 
$m_\varphi = 120$ GeV/c$^2$ these cuts correspond to
48, 34 and 15 GeV for the leading jet, second and third leading jets
of the event, respectively, and vary roughly linearly with the Higgs 
boson mass.
We then require that at least three among the four highest-$E_T$
jets in the event are identified (tagged) as $b$ quark candidates.
The algorithm used to identify secondary vertices~\cite{secvtx}
begins by searching for jets which contain three or more displaced tracks.
If none are found, the algorithm
searches for two-track vertices using more stringent track criteria. A jet
is tagged if the transverse displacement of the secondary vertex 
from the primary vertex exceeds three times its uncertainty.
A requirement on the azimuthal angular distribution of the two highest-$E_T$ 
$b$-tagged jets in the event, $\Delta\varphi_{b\bar{b}}>109^{\circ}$,
reduces the heavy flavor QCD content of the sample attributed 
to gluon splitting.
This cut preserves $\sim 90$\% of the signal events
which favor a larger angular separation between the $b$-tagged
jets coming from the Higgs decay.
After the three tag requirement and the $\Delta\varphi$ cut we are 
left with 20 and 13 events, respectively, for the case of the
$m_\varphi = 70$ GeV/c$^2$ selection.
\par
To reconstruct the signal we select one of the possible invariant mass
combinations bet\-ween the highest-$E_T$ jets in the event. 
From Monte Carlo we find that the mass of the highest-$E_T$ jets in the event
($m_{12}$) for signal masses $m_\varphi>120$ GeV/c$^2$, 
and the mass of the second and third highest-$E_T$ jets 
in the event ($m_{23}$) for signal masses $m_\varphi\le 120$ GeV/c$^2$
enhance the signal resolution, $\delta(m_{\varphi})/m_\varphi$.
The use of these distributions also minimizes
the percentage of events for which at least one of the
jets in the dijet mass is not associated with a $b$ quark from 
a Higgs boson decay. All signal mass distributions contain
a Gaussian core with a resolution which depends on the Higgs boson mass 
and varies from $\sim$25\% for $m_\varphi\le 120$ GeV/c$^2$ to 
$\sim$13\% for $m_\varphi> 120$ GeV/c$^2$. 
We increase the expected significance of the signal by applying mass window 
cuts on the $m_{12}$ and $m_{23}$ distributions 
which vary between $\pm 1\delta(m_{\varphi})$ and 
$\pm 3\delta(m_{\varphi})$, depending on the Higgs mass, and centered 
on the mean of the fit distributions. A cut on the 
invariant mass distribution between the
two $b$-tagged highest-$E_T$ jets of the event further discriminates 
against heavy flavor QCD events.
All mass cuts were chosen to maximize the expected significance
of the signal.
Table~\ref{tab1} shows the number of observed 
triple $b$-tagged events left after all cuts as a function of mass.
Five events are left after all cuts in the mass bin at 70 GeV/c$^2$. 
All events for the mass bins above 70 GeV/c$^2$ are included
in this sample of five events.
\par
In addition to the large QCD multijet background, other sources of 
heavy flavor in the triple $b$-tagged sample include multijet 
$t\bar{t}$ production ($t\rightarrow Wb$, $W\rightarrow q\bar{q}'$),
$Wb\bar{b}$ and $Wc\bar{c}$ with $W\rightarrow q\bar{q}'$,
$Zb\bar{b}$ and $Zc\bar{c}$ with $Z\rightarrow b\bar{b}/c\bar{c}$, 
and fake triple-tags. They are estimated
from a combination of Monte Carlo and data. 
We use the HERWIG v5.6 Monte Carlo generator~\cite{herwig} with
the CDF measured cross section 
($\sigma_{t\bar{t}} = 6.5^{+1.7}_{-1.4}$ pb)~\cite{pisa} and a top mass 
of $m_t = 175$ GeV/$c^2$ to predict the expected number of $t\bar{t}$ 
events. Electroweak processes are also estimated with the same Monte Carlo
generator. Fake triple-tags are defined as events 
in which at least one of the three $b$-tagged jets contains a false 
secondary vertex in a light quark or gluon jet. Fake tag probabilities 
are parameterized by measuring in several inclusive jet data samples the 
fraction of jets in which a secondary vertex is reconstructed on the wrong
side of the primary vertex with respect to the jet 
direction~\cite{secvtx,top_paper}.

\begin{figure}[p]
\epsfxsize=6.0in
\epsfysize=6.0in
\centerline{\hspace{-0.1cm}\epsffile{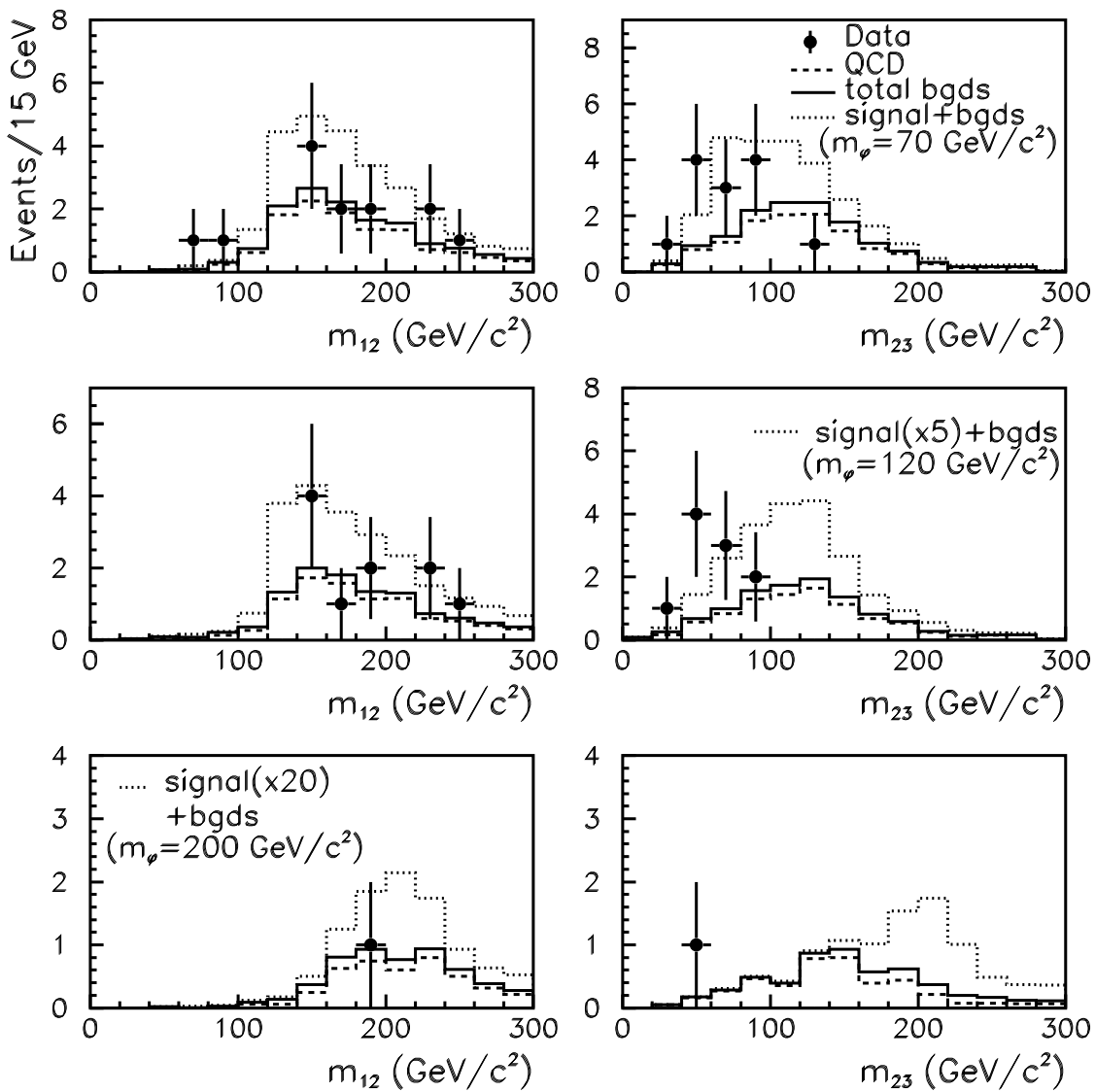}}
\vspace*{0.4cm}
\caption{Invariant mass distributions $m_{12}$ and
$m_{23}$ for the observed triple $b$-tagged sample co\-rres\-pon\-ding
to the $m_{\varphi}=70$ GeV/c$^2$ (top),
$m_{\varphi}=120$ GeV/c$^2$ (middle), and $m_{\varphi}=200$ GeV/c$^2$
(bottom) selections. The data is compared to the expected 
QCD only background, the total SM backgrounds, and the total 
background plus signal for $\tb = 50$ and the {\em no mixing} case. 
The use of $m_{23}$ for $m_\varphi \le 120$ GeV/c$^2$, and $m_{12}$ for
$m_\varphi > 120$ GeV/c$^2$ maximizes the fraction of correct jet 
assignments and enhance the signal resolution (see text). 
The mass cuts are not applied.}
\label{fig1}
\end{figure}

The same fit technique that was used to estimate the QCD 
heavy flavor normalization in the SM Higgs search~\cite{cdfhad}
is applied to this analysis.
We reduce the $b$-tag cuts on our data sample from triple to double
$b$-tags. This gives a high statistics background-rich sample
in which we fit the double $b$-tagged dijet mass distribution
to a combination of signal and SM backgrounds.
The shape of the QCD heavy flavor distribution is obtained from the 
PYTHIA Monte Carlo program.
We generate all QCD jet production channels,
and retain the events that contain a heavy quark produced either
in the hard scattering or in the associated radiation process.
The signal and QCD normalization is left free in the fit while
the SM non-QCD background, both shape and normalization, is obtained
from Monte Carlo simulation. 
The QCD normalization in our triple $b$-tagged sample is then obtained
from the ratio of normalizations of double $b$-tagged
to triple $b$-tagged QCD events taken
from the PYTHIA Monte Carlo simulation.
Table~\ref{tab2} lists the individual QCD, $t\bar{t}$, 
$Wb\bar{b}$, $Wc\bar{c}$, $Zb\bar{b}$, $Zc\bar{c}$, fake triple tags,
and total expected contributions to the final observed sample as a 
function of mass. From these numbers we find no evidence for the presence 
of a Higgs boson signal. Figure~\ref{fig1} shows the $m_{12}$ and $m_{23}$
distributions for the observed triple $b$-tagged sample compared to the SM 
background expectations and for three different
selections corresponding to $m_\varphi = 70$, $120$, and $200$ GeV/c$^2$.  
Also shown are the signal plus background shapes normalized to the
expected number of events for $\tb = 50$ and the case of no mixing in
the scalar top sector ({\em no mixing} scenario). 
\par
We calculate the signal detection efficiencies and normalizations
from Monte Carlo.
The overall detection acceptances with their total uncertainties
are shown in Table~\ref{tab1} as a function of mass. 
They are within a range of 0.2\% to 0.6\%, increasing with mass.
This low acceptance is dominated by the small multijet trigger efficiencies
($\sim$1\% to $\sim$7\%, increasing with the signal mass) and, 
to a lesser extent, by the triple $b$-tag requirement ($\sim$20\%).
The low values for the trigger efficiency are due to the high
$E_T$ thresholds and multiplicity requirements on jets. The trigger efficiency
curves have been obtained with a trigger simulation with parametrized
curves estimated from data.
The total systematic error includes uncertainties
in the multijet trigger simulation ($5$\%), in the modelling of gluon 
radiation ($10$\% to $7$\%, depending on the mass), in the calorimeter 
energy scale ($10$\% to $2$\%, depending on the mass), in the luminosity 
measurement ($7$\%), and in the $b$-tag efficiencies ($10$\%).
\par
\begin{figure}[p]
\epsfxsize=6.2in
\epsfysize=6.2in
\centerline{\hspace{-0.1cm}\epsffile{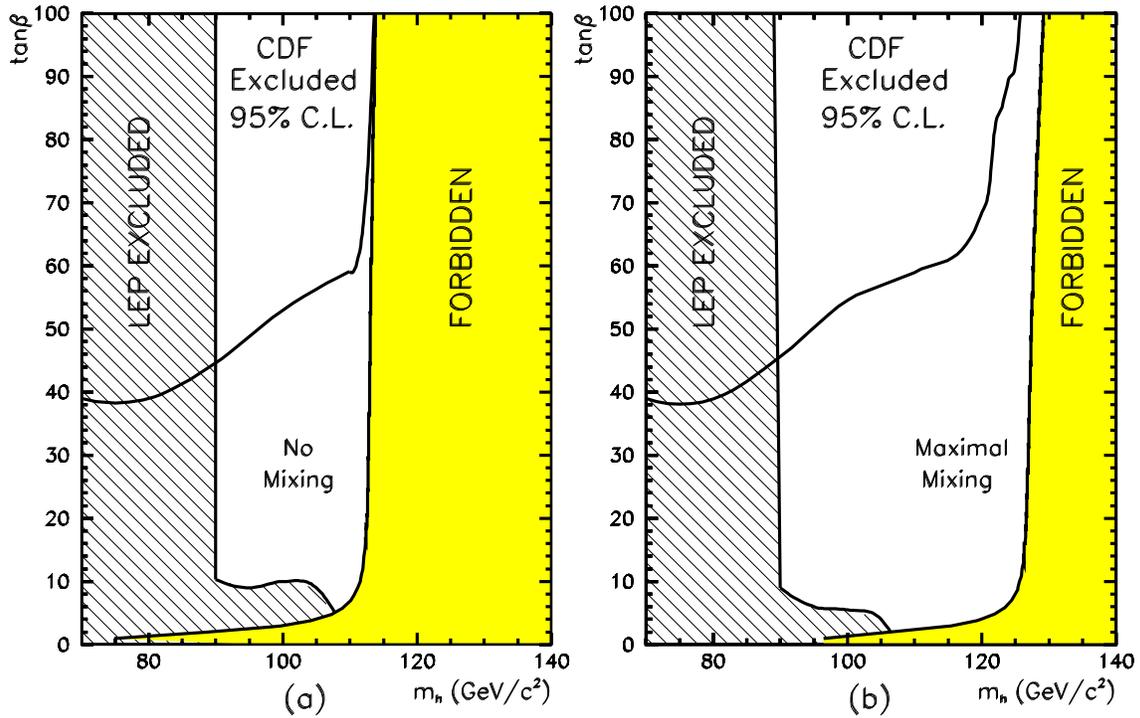}}
\vspace*{0.3cm}
\caption{CDF 95\% C.L. excluded region in the parameter space
$m_h-\tb$ for the two stop mixing scenarios: (a) {\em no mixing},
and (b) {\em maximal mixing}. Also shown are the theoretically forbidden 
regions and the LEP exclusion region for their {\em no mixing} and
{\em $m_h^{max}$} scenarios~[16].}
\label{fig2}
\end{figure}
\begin{figure}[p]
\epsfxsize=6.2in
\epsfysize=6.2in
\centerline{\hspace{-0.1cm}\epsffile{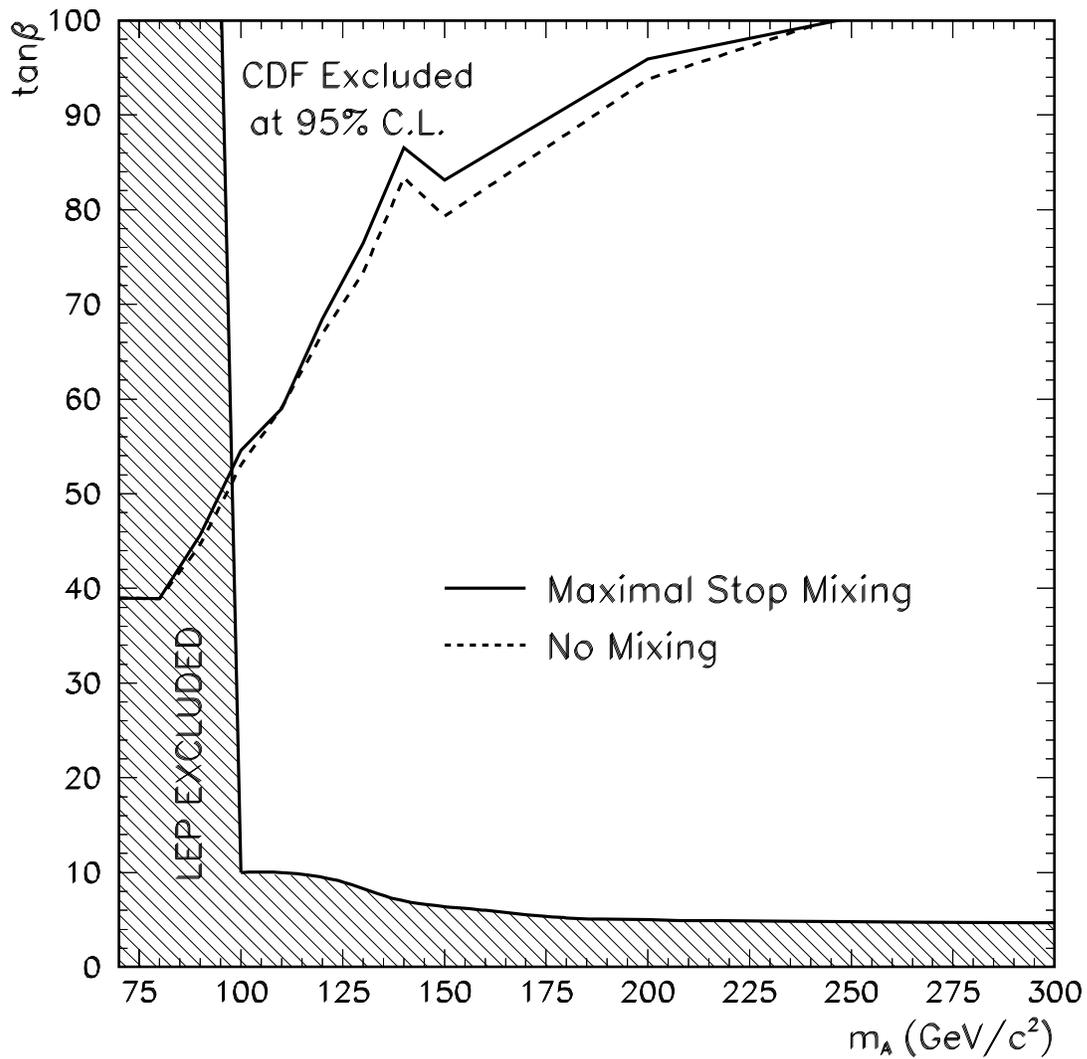}}
\vspace*{0.3cm}
\caption{CDF 95\% C.L. excluded region in the parameter space
$m_A-\tb$ for the two stop mixing scenarios: {\em no mixing} 
(dashed lines) and {\em maximal mixing} (solid line). Also shown is
the LEP exclusion region for the {\em no mixing} scenario~[16].}
\label{fig3}
\end{figure}

From the data in Table~\ref{tab1} we set upper limits 
on $b\bar{b}\varphi\rightarrow b\bar{b}b\bar{b}$
($\varphi = h,H,A$) production using a one-sided conditional
frequentist construction~\cite{zech}, where 
systematic uncertainties are approximately taken into
account by Bayesian averaging over the systematic parameters~\cite{gp},
assuming gaussian a-priori distributions around their best estimates.
The 95\% C.L. upper limits on the total expected signal events as well
as on the production cross section times branching fraction are listed 
in Table~\ref{tab1}. Using the LO theoretical cross sections for
$\sigma(p\bar{p}\rightarrow b\bar{b}\varphi) 
{\cal BR}(\varphi\rightarrow b\bar{b})$
with $\varphi = h,H,A$ and the bottom-Higgs Yukawa coupling calculated
with a running bottom quark mass evaluated at the Higgs boson mass scale,
$m_b(m_\varphi)\simeq 3$ GeV/c$^2$,
we exclude regions of parameter space 
for $m_h - \tb$ and $m_A - \tb$, respectively, as shown in 
Figures~\ref{fig2} and \ref{fig3}. Results are shown 
for two common choices of the stop quark mixing parameter~\cite{fin}:
{\em no  mixing} ($A_t=\mu\cot\beta$), and {\em maximal mixing} 
($A_t=\mu\cot\beta + \sqrt{6}\,m_S$, with $\mu$ the supersymmetric
Higgs boson mass parameter and $A_t$ a soft breaking parameter). In all
cases we set $m_S$, the supersymmetric mass scale,
to be 1 TeV/c$^2$ and $m_{t} = 175$ GeV/c$^2$.
As a test of the sensitivity of our calculated limits to the background
estimate, if we assume a zero background hypothesis the result limits
increase, for all signal masses, by $\delta(\tan\beta)=10$.
\par
In conclusion, we have searched for neutral Higgs bosons produced in 
association with $b$ quarks through the reaction 
$p\bar{p}\rightarrow b\bar{b} \varphi\rightarrow b\bar{b}b\bar{b}$.
We do not find evidence for the presence of a signal and 
95\% C.L. upper limits are set on the production cross section times
branching ratio. The results have been interpreted in the
context of the MSSM Higgs sector and lower mass limits for neutral Higgs
bosons derived for $\tan\beta$ values in excess of 35.
\par
We thank the Fermilab staff and the technical staffs of the
participating institutions for their vital contributions. 
This work was supported
by the U.S. Department of Energy, the National Science Foundation,
the Istituto Nazionale di Fisica Nucleare (Italy), 
the Ministry of Science, Culture and Education of Japan, 
the Natural Sciences and Engineering Council of Canada, 
the National Science Council of the Republic of China,
and the A.~P.~Sloan Foundation.

\renewcommand{\baselinestretch}{1.0}

\begin{table}[h]
\begin{center}
\caption{Number of observed and expected background events
after the final selection and
for the different SM contributions as a function of mass. 
Last three columns show the total acceptances and expected 
95\% C.L. upper limits on the number of signal events ($N_{signal}$) and on
$\sigma\times {\cal BR}$, respectively.}
\label{tab1}
\begin{tabular}{cccccc}
$m_\varphi$  & Observed & Expected & 
Acceptance & $N_{signal}$  & $\sigma\times {\cal BR}$  \\
(GeV/c$^2$) & Events & Background & (\%)& (95\% C.L.) & 
(pb, 95\% C.L.) \\ [0.1cm]\hline
70  & 5 & $4.6 \pm 1.4$ & $0.16 \pm 0.03$ & $7.9$ & $53.3$  \\
80  & 4 & $4.6 \pm 1.4$ & $0.22 \pm 0.04$ & $6.6$ & $31.7$ \\
90  & 3 & $3.8 \pm 1.3$ & $0.23 \pm 0.04$ & $5.8$ & $27.1$ \\
100 & 3 & $3.8 \pm 1.3$ & $0.25 \pm 0.04$ & $5.9$ & $25.7$ \\
110 & 2 & $3.7 \pm 1.1$ & $0.25 \pm 0.04$ & $4.8$ & $20.7$ \\
120 & 2 & $3.5 \pm 1.1$ & $0.28 \pm 0.05$ & $4.9$ & $19.2$ \\
130 & 1 & $2.6 \pm 0.9$ & $0.28 \pm 0.05$ & $4.1$ & $15.8$ \\
140 & 1 & $1.7 \pm 0.8$ & $0.29 \pm 0.05$ & $4.3$ & $16.2$ \\
150 & 0 & $1.5 \pm 0.8$ & $0.30 \pm 0.05$ & $3.2$ & $11.5$ \\
200 & 0 & $1.2 \pm 0.7$ & $0.41 \pm 0.07$ & $3.2$ & $ 8.5$ \\
250 & 0 & $1.0 \pm 0.7$ & $0.47 \pm 0.08$ & $3.2$ & $ 7.5$ \\
300 & 0 & $0.1 \pm 0.4$ & $0.59 \pm 0.09$ & $3.2$ & $ 5.8$ 
\end{tabular}
\end{center}
\end{table} 

\begin{table}[h]
\begin{center}
\caption{Expected QCD, fake triple tags, $t\bar{t}$,
$Wb\bar{b} + Wc\bar{c}$, $Zb\bar{b} + Zc\bar{c}$, and total
number of background events as a function of mass.}
\label{tab2}
\begin{tabular}{ccccccc}
$m_\varphi$ (GeV/c$^2$) & QCD & Fakes & $t\bar{t}$ & $Wb\bar{b} + Wc\bar{c}$
& $Zb\bar{b} + Zc\bar{c}$ & Total \\ [0.1cm]\hline
70  & $2.97 \pm 0.70$ & $0.5 \pm 1.2$ & $0.70 \pm 0.18$ & $0.09 \pm 0.06$ & 
$0.37 \pm 0.02$ & $4.6 \pm 1.4$  \\
80  & $2.97 \pm 0.70$ & $0.5 \pm 1.2$ & $0.70 \pm 0.18$ & $0.09 \pm 0.06$ & 
$0.37 \pm 0.02$ & $4.6 \pm 1.4$ \\
90  & $2.16 \pm 0.55$ & $0.5 \pm 1.2$ & $0.70 \pm 0.18$ & $0.09 \pm 0.06$ & 
$0.37 \pm 0.02$ & $3.8 \pm 1.3$ \\
100 & $2.16 \pm 0.55$ & $0.5 \pm 1.2$ & $0.70 \pm 0.18$ & $0.09 \pm 0.07$ & 
$0.37 \pm 0.02$ & $3.8 \pm 1.3$ \\
110 & $2.16 \pm 0.55$ & $0.4 \pm 0.9$ & $0.68 \pm 0.18$ & $0.09 \pm 0.07$ & 
$0.37 \pm 0.02$ & $3.7 \pm 1.1$ \\
120 & $2.16 \pm 0.55$ & $0.3 \pm 0.9$ & $0.66 \pm 0.17$ & $0.07 \pm 0.06$ & 
$0.29 \pm 0.02$ & $3.5 \pm 1.1$ \\
130 & $1.44 \pm 0.45$ & $0.3 \pm 0.8$ & $0.64 \pm 0.17$ & $0.05 \pm 0.05$ & 
$0.21 \pm 0.02$ & $2.6 \pm 0.9$ \\
140 & $0.73 \pm 0.40$ & $0.2 \pm 0.7$ & $0.60 \pm 0.16$ & $0.04 \pm 0.05$ & 
$0.16 \pm 0.02$ & $1.7 \pm 0.8$ \\
150 & $0.73 \pm 0.40$ & $0.1 \pm 0.7$ & $0.55 \pm 0.15$ & $0.03 \pm 0.05$ & 
$0.12 \pm 0.02$ & $1.5 \pm 0.8$ \\
200 & $0.73 \pm 0.40$ & $0.1 \pm 0.6$ & $0.32 \pm 0.09$ & $0.00 \pm 0.00$ & 
$0.00 \pm 0.00$ & $1.2 \pm 0.7$ \\
250 & $0.73 \pm 0.40$ & $0.1 \pm 0.6$ & $0.17 \pm 0.05$ & $0.00 \pm 0.00$ & 
$0.00 \pm 0.00$ & $1.0 \pm 0.7$ \\
300 & $0.01 \pm 0.01$ & $0.1 \pm 0.4$ & $0.00 \pm 0.00$ & $0.00 \pm 0.00$ & 
$0.00 \pm 0.00$ & $0.1 \pm 0.4$ \\
\end{tabular}
\end{center}
\end{table} 

\end{document}

%% file: authors.tex
\font\eightit=cmti8
\def\r#1{\ignorespaces $^{#1}$}
\hfilneg
\begin{sloppypar}
\noindent
T.~Affolder,\r {23} H.~Akimoto,\r {45}
A.~Akopian,\r {38} M.~G.~Albrow,\r {11} P.~Amaral,\r 8 S.~R.~Amendolia,\r {34} 
D.~Amidei,\r {26} K.~Anikeev,\r {24} J.~Antos,\r 1 
G.~Apollinari,\r {11}
% ...
\end{sloppypar}
\vskip .026in
\begin{center}
(CDF Collaboration)
\end{center}

\vskip .026in
\begin{center}
\r 1  {\eightit Institute of Physics, Academia Sinica, Taipei, Taiwan 11529, 
Republic of China} \\
\r 2  {\eightit Argonne National Laboratory, Argonne, Illinois 60439} \\
\r 3  {\eightit Istituto Nazionale di Fisica Nucleare, University of Bologna,
I-40127 Bologna, Italy} \\
\r 4  {\eightit Brandeis University, Waltham, Massachusetts 02254} \\
\r 5  {\eightit University of California at Davis, Davis, California  95616} \\
\r 6  {\eightit University of California at Los Angeles, Los 
Angeles, California  90024} \\  
\r 7  {\eightit Instituto de Fisica de Cantabria, CSIC-University of Cantabria, 
39005 Santander, Spain} \\
\r 8  {\eightit Enrico Fermi Institute, University of Chicago, Chicago, 
Illinois 60637} \\
\r 9  {\eightit Joint Institute for Nuclear Research, RU-141980 Dubna, Russia}
\\
\r {10} {\eightit Duke University, Durham, North Carolina  27708} \\
\r {11} {\eightit Fermi National Accelerator Laboratory, Batavia, Illinois 
60510} \\
\r {12} {\eightit University of Florida, Gainesville, Florida  32611} \\
\r {13} {\eightit Laboratori Nazionali di Frascati, Istituto Nazionale di Fisica
               Nucleare, I-00044 Frascati, Italy} \\
\r {14} {\eightit University of Geneva, CH-1211 Geneva 4, Switzerland} \\
\r {15} {\eightit Glasgow University, Glasgow G12 8QQ, United Kingdom}\\
\r {16} {\eightit Harvard University, Cambridge, Massachusetts 02138} \\
\r {17} {\eightit Hiroshima University, Higashi-Hiroshima 724, Japan} \\
\r {18} {\eightit University of Illinois, Urbana, Illinois 61801} \\
\r {19} {\eightit The Johns Hopkins University, Baltimore, Maryland 21218} \\
\r {20} {\eightit Institut f\"{u}r Experimentelle Kernphysik, 
Universit\"{a}t Karlsruhe, 76128 Karlsruhe, Germany} \\
\r {21} {\eightit Center for High Energy Physics: Kyungpook National
University, Taegu 702-701; Seoul National University, Seoul 151-742; and
SungKyunKwan University, Suwon 440-746; Korea} \\
\r {22} {\eightit High Energy Accelerator Research Organization (KEK), Tsukuba, 
Ibaraki 305, Japan} \\
\r {23} {\eightit Ernest Orlando Lawrence Berkeley National Laboratory, 
Berkeley, California 94720} \\
\r {24} {\eightit Massachusetts Institute of Technology, Cambridge,
Massachusetts  02139} \\   
\r {25} {\eightit Institute of Particle Physics: McGill University, Montreal 
H3A 2T8; and University of Toronto, Toronto M5S 1A7; Canada} \\
\r {26} {\eightit University of Michigan, Ann Arbor, Michigan 48109} \\
\r {27} {\eightit Michigan State University, East Lansing, Michigan  48824} \\
\r {28} {\eightit University of New Mexico, Albuquerque, New Mexico 87131} \\
\r {29} {\eightit The Ohio State University, Columbus, Ohio  43210} \\
\r {30} {\eightit Osaka City University, Osaka 588, Japan} \\
\r {31} {\eightit University of Oxford, Oxford OX1 3RH, United Kingdom} \\
\r {32} {\eightit Universita di Padova, Istituto Nazionale di Fisica 
          Nucleare, Sezione di Padova, I-35131 Padova, Italy} \\
\r {33} {\eightit University of Pennsylvania, Philadelphia, 
        Pennsylvania 19104} \\   
\r {34} {\eightit Istituto Nazionale di Fisica Nucleare, University and Scuola
               Normale Superiore of Pisa, I-56100 Pisa, Italy} \\
\r {35} {\eightit University of Pittsburgh, Pittsburgh, Pennsylvania 15260} \\
\r {36} {\eightit Purdue University, West Lafayette, Indiana 47907} \\
\r {37} {\eightit University of Rochester, Rochester, New York 14627} \\
\r {38} {\eightit Rockefeller University, New York, New York 10021} \\
\r {39} {\eightit Rutgers University, Piscataway, New Jersey 08855} \\
\r {40} {\eightit Texas A\&M University, College Station, Texas 77843} \\
\r {41} {\eightit Texas Tech University, Lubbock, Texas 79409} \\
\r {42} {\eightit Istituto Nazionale di Fisica Nucleare, University of Trieste/
Udine, Italy} \\
\r {43} {\eightit University of Tsukuba, Tsukuba, Ibaraki 305, Japan} \\
\r {44} {\eightit Tufts University, Medford, Massachusetts 02155} \\
\r {45} {\eightit Waseda University, Tokyo 169, Japan} \\
\r {46} {\eightit University of Wisconsin, Madison, Wisconsin 53706} \\
\r {47} {\eightit Yale University, New Haven, Connecticut 06520} \\
\r {(\ast)} {\eightit Now at Carnegie Mellon University, Pittsburgh,
Pennsylvania  15213}
\end{center}

%% file: mssm_sent_to_prl_v10.bbl
\begin{references}

\bibitem{fin}
M. Carena {et al.}, CERN-TH-99-374, hep-ph/9912223.

\bibitem{cdflep}
F. Abe {\it et al.}, Phys. Rev. Lett. {\bf 79}, 3819 (1997).

\bibitem{cdfhad}
F. Abe {\it et al.}, Phys. Rev. Lett. {\bf 81}, 5748 (1998).

\bibitem{cdf_d}
F. Abe {\it et al.}, Nucl. Instrum. Methods Phys. Res., Sect. A 
{\bf 271}, 387 (1988).

\bibitem{svx}
D. Amidei {\it et al.}, Nucl. Instrum. Methods Phys. Res., Sect. A 
{\bf 350}, 73 (1994).

\bibitem{CDFcoo} 
In the CDF coordinate system, $\phi$ and $\theta$ are the
azimuthal and polar angles with respect to
the proton beam direction.
The pseudorapidity $\eta$
is defined as $-\ln[\tan (\theta/2)]$.
The transverse momentum of a particle
is $\pt = p \sin \theta$, where $p$ is the momentum
measured in the spectrometer.
The analogous quantity using calorimeter energies is
called the transverse energy $\et$.
The difference between the vector sum of all the
transverse energies and zero is the missing transverse energy $\met$.

\bibitem{jets}
F. Abe {\it et al.}, 
Phys. Rev. D {\bf 45}, 1448 (1992).

\bibitem{tim}
C. Bal{\'a}zs, J. L. Diaz-Cruz, H.-J. He, T. Tait, and C.-Y. Yuan,
Phys. Rev. D {\bf 59}:055016 (1999);
M. Carena, S. Mrenna, C.E.M. Wagner, Phys. Rev.D {\bf 60}:075010 (1999). 

\bibitem{pythia}
T. Sj\"{o}strand, Comput. Phys. Commun. {\bf 82}, 74 (1994).

\bibitem{secvtx}
F. Abe {\it et al.}, 
Phys. Rev. Lett. {\bf 74}, 2626 (1995).

\bibitem{herwig}
G. Marchesini {\it et al.}, Comput. Phys. Commun. {\bf 67}, 465 (1992).

\bibitem{pisa}
F. Ptohos (for the CDF collaboration), proceedings of the International
Europhysics Conference on High Energy Physics 99, Tampere, Finland, July 17,
1999.

\bibitem{top_paper}
F. Abe {\it et al.}, 
Phys. Rev. Lett. {\bf 79}, 1992 (1997).

\bibitem{zech}
G. Zech, Nucl. Instrum. Methods Phys. Res., Sect A{\bf 277}, 608 (1989).
This formula also admits a Bayesian interpretation, see O. Helene,
Nucl. Instrum. Methods, {\bf 212}, 319 (1983). See also the discussion
of eq. (7) in B. P. Roe and M. B. Woodroofe, Phys. Rev. {\bf D60} 053009 
(1999).

\bibitem{gp}
See for instance R. Cousins and V. Highland, Nucl. Instrum. Methods
Phys. Res., Sect A{\bf 320}, 331 (1992). We use numerical techniques
to solve the general case.

\bibitem{lepwg}
The LEP working group for Higgs boson searches. CERN-EP/2000-055;
ALEPH 2000-028 CONF 2000-023;
DELPHI 2000-050 CONF 365; L3 note 2525; OPAL TN646, March 15, 2000.

\end{references}
